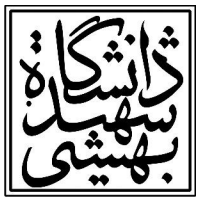

# Design Flow of Digital Microfluidic Biochips

# Towards Improving Fault-Tolerance

by

Alireza Abdoli

A thesis submitted in partial fulfillment of
the requirements for the degree of

Master of Science

Shahid Beheshti University

2015

SHAHID BEHESHTI UNIVERSITY

ABSTRACT

**Design Flow of Digital Microfluidic Biochips Towards**

**Improving Fault-Tolerance**

by Alireza Abdoli

Chairperson of the Supervisory Committee:

Professor Ali Jahanian

Department of Computer Science and Engineering

Given the ever-increasing advances of digital microfluidic biochips and their application in a wide range of areas including bio-chemistry experiments, diagnostics, and monitoring purposes like air and water quality control and etc., development of automated design flow algorithms for digital microfluidic biochips is of great importance. During the course of last decade there have been numerous researches on design, adaptation and optimization of algorithms for automation of digital microfluidic biochips synthesis flow.

However, the initial assumption of researchers about absence of faults and deficiencies before and during execution of bio-assays has been proven always not to be the case. Thus, during the past few years researchers have placed great focus on fault-tolerance and fault-recovery of digital microfluidic biochips.

In this dissertation we initially introduce proposed architectures for pin-constrained digital microfluidic biochips; the proposed architectures are designed with the aim of improving overall functionality and also at the same time ameliorating fault-tolerance of digital microfluidic biochips in mind.

The proposed architectures in this dissertation include general-purpose field-programmable pin-constrained architecture and field-programmable cell array pin-constrained architecture. Comparing the general-purpose field-programmable pin-constrained architecture versus the base architecture dimension of the array of electrodes is reduced by 33%. Also, regarding number of electrodes and controlling pins 20% and 3% reductions are observed, respectively. Both architectures provide similar performance in terms of microfluidic operation times. However, droplet routing times are reduced by 17%. Finally, given the aforementioned factors 2% reduction is achieved in total times.

Considering the base architecture versus the field-programmable cell array architecture dimension of array of electrodes is reduced by 10%. Regarding the number of electrodes and controlling pins 1% and 8% reductions are observed, respectively. Both architectures provide similar performance in terms of microfluidic operation times. Regarding droplet routing time 12% reduction, versus the base architecture, is achieved. Finally, considering the total bioassay execution time 1% reduction is achieved.

Next, we explain fault-tolerance concepts within the context of pin-constrained digital microfluidic biochips; then we attempt to investigate fault-tolerance of the proposed digital microfluidic architectures versus the base architecture in presence of faults occurrences affecting mixing modules and Split / Storage / Detection (SSD) modules.

*Chapter 1*

# MICROFLUIDICS

Over the past decade, microfluidics as a valuable technology has played an important role in automating and minimizing biochemical processes. Microfluidic devices allow us to use much less fluid (on a volumetric scale) than conventional fluids in milliliters, as well as conventional laboratory devices, and there is no need for conventional laboratory devices. Residual technology enables us to perform many chemical experiments on laboratory devices on a chip (i.e. Lab-On-Chip (LoC). Laboratory devices on a chip perform a variety of biochemical functions such as diagnostic tests (In-Vitro) and immunoassays, DNA polymerase chain reaction (PCR) [1].

This thesis addresses digital microfluidic biochips (DMFB), in which the droplets are controlled individually and simultaneously on a two-dimensional array of cells (electrodes). A major advantage of digital microfluidic biochips is their reconfigurability, ease of integration and scalability [2].

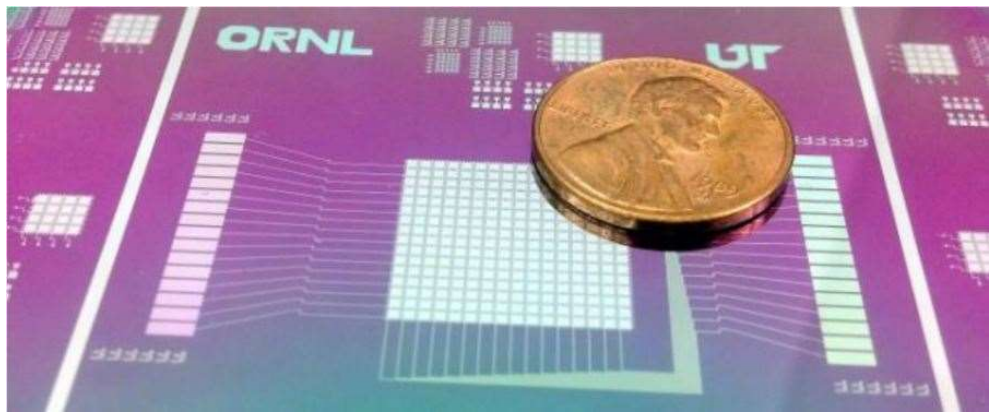

Figure 1. A typical DMFB device [1]

The benefits of digital microfluidic biochips are as follows [3]:

- Absence of mechanical components: All recycling operations are carried out between the upper and lower two layers of the bio-chip.

- No need for channels: The gap between the two plates is filled by the filler fluid.

- Ability to independently control droplets: to apply localized electric humidification force.

- Controlling or preventing evaporation: The filler (oil) surrounding the droplets prevents evaporation.

- Non-use of electric current: Despite capacitive currents, it does not occur to avoid direct sample heating and electrochemical reactions.

- Compatibility with a wide variety of liquids: Most electrolyte solutions.

- Near-full productivity of the sample or reactor: due to no loss of fluid to cover the ducts or fill the tanks.

- Compatibility for Microscopic Applications: Glass substrate and Indium Tin Oxide (ITO) electrodes allow microscopic examination.

- High Energy Efficiency: Each drop transfer consumes energy at the nanoscale or microwave scale.

- High droplet speeds: up to 25 centimeters per second.

- Using Protocols Based on Equivalent Laboratory Liquid Chemistry: Users can scale, automate and embed proven experiments and protocols.

- Maximum operational flexibility: Direct computer control at each step enables conditional execution of the next steps.

As shown in Fig. (Fig: DMFB), the digital microfluidic biochip is composed of two plates coated with a hydrophobic layer of Teflon Amorphous Fluoroplastics. The bottom plate consists of an array of control electrodes, while the top plate consists of a single conductive electrode that covers the entire top plate [1]. The liquid droplets move between the top and bottom plates and into the fluid filling the space between the top and bottom plates.

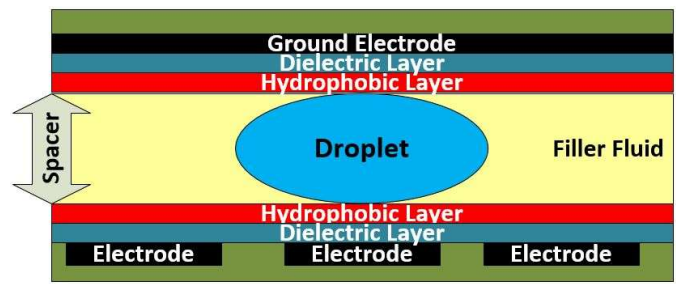

Figure 2. DMFB Structure

All electrodes are capable of performing general microfluidic operations such as compound storage, splitting, etc., although some electrodes are equipped with special modules such as optical or heat sensors to detect and heat recrystallization if needed. As mentioned, all electrodes are capable of performing general microfluidic operations, so the electrodes can be used to perform general microfluidic operations if there is no need to use special modules embedded in digital microfluidic bio-chips (such as sensors, heaters, etc.).

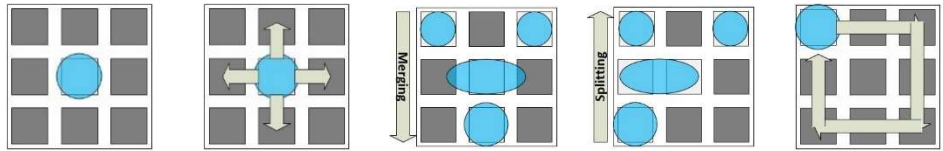

Figure 3. Microfluidic operations from left to right; storage, transporting, merging, splitting and mixing [1]

The first microfluidic operation is to keep the drop constant, by applying voltage to the electrode beneath the drop. This is important because all the droplets on the biochip must be kept in place during storage on the biochip, otherwise the droplet may slip and the process may be disrupted.

The second microfluidic operation is droplet transfer which occurs by applying voltage to one of the adjacent electrodes and simultaneously deactivating the electrode below the droplet.

The third microfluidic operation is to merge the two droplets together. In this operation, two droplets are moved adjacent to each other and then the electrode beneath one of the droplets is deactivated while the electrode is still active beneath the other droplet.

The fourth microfluidic operation is to divide a drop into two droplets of approximately identical volumes; this involves activating electrodes adjacent to the droplet (left and right electrodes, or top and bottom electrodes) and simultaneously disabling the electrode beneath the droplet. Ideally, the large droplet will be split into two droplets with approximately equal volumes due to tendency to move towards the active electrodes.

The fifth microfluidic operation is to combine two droplets to obtain a uniform and homogeneous combination of the two droplets; first, by using the merging operation, the two droplets are combined, then moved the droplet on a specified path of electrodes for a specified period of time. The movement is to obtain a uniform and homogeneous droplet.

Synthesis of digital microfluidic biochips consists of three primary stages: Scheduling, Placement, and Droplet Routing, and two preliminary stages of Pin-Mapping and Pin Routing [11]. The following describes the structure of a digital microfluidic system and each step of the automation process for the physical design of a digital microfluidic biochip.

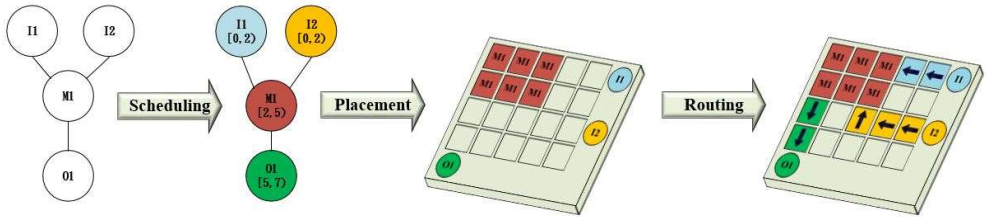

Figure 4. Synthesis flow of DMFB devices [4] [5] [6]

Scheduling is the first step in the synthesis of DMFBs. Each microfluidic operation is given an exact start and end time. The performance of the scheduling algorithm must be such as to ensure that no operations are initiated before the end of its parent operations and that there are sufficient resources available to execute the operations that are scheduled simultaneously [1].

After the scheduling phase is over, it is time to start the placement phase. The placement phase decides where to execute microfluidic operations on the surface of the DMFB device [1].

After the timing of the operations and the placement of the modules on the surface of the 2D array of electrodes, it is time for the droplet routing step. At this stage, the routing operation of the liquid droplets is performed to move between the input reservoirs and the modules, between the modules and the movement from the modules to the output reservoirs. At the routing stage, it must be ensured that the droplet reaches its designated destination and does not interfere with other droplets [1].

After the droplet routing step is completed, an optional step is sometimes pin-mapping. Digital microfluidic biochips pin-mapping schemes can be categorized as individually addressable pins (sometimes referred to as direct addressing), pin-constrained and active-matrix.

Finally, the last step in the synthesis is the routing of the wires. The electrodes are located on the bottom plate of the digital microfluidic bio-chip, and routing the wires is done beneath this substrate onto one or more layers of the printed circuit board (PCB) or other material including glass [7] or paper [8] [9].

Proposed solutions include pin-constrained DMFB architectures with the aim of outperforming previous architectures while improving the fault tolerance of the proposed architectures.

Here are some overviews of suggested solutions:

**Time Overhead:** We will have to use other modules due to permanent hardware faults in modules, which in most cases will result in increased operation time and droplet routing time.

**Space Overhead:** Other modules should be used if possible, in order to avoid faulty modules.

The following can be mentioned in terms of improvements to the proposed DMFB architectures:

- Improvements in DMFB hardware (reduction of DMFB dimensions, reduction of the number of electrodes and control pins used to drive the DMFB device)

- Significantly decreasing the droplet routing time that the overall execution time of bioassays is decreased.

- Adding hardware/software fault-tolerance capability so that even in the event of destructive (permanent) faults, the DMFB can continue to operate and perform biological tests.

*Chapter 2*

PROPOSED DMFB ARCHITECTURES

In this section, we introduce and evaluate the performance of fault-tolerant digital microfluidic biochip architectures. In the process of designing the proposed architectures, special attention has been paid to improving fault tolerance. We will first introduce the benchmark architecture, which will be used to compare the performance of the proposed architectures in this chapter.

Before presenting the proposed architectures, we introduce a set of commonly used bioassays along with their characteristics. We then introduce the proposed digital biochip architectures and examine the performance of the architectures assuming no faults and faults. Finally, we describe the addition and improvements on fault tolerance in digital pin-constrained biochips. Adding fault tolerance capability to this chapter is associated with the mixing modules along with the splitting/storing/detecting modules.

- PCR or polymerase chain reaction is a method of amplifying small amounts of DNA or RNA for use in laboratory applications. Digital microfluidic biochips are used to perform a set of recursive PCR experiments.

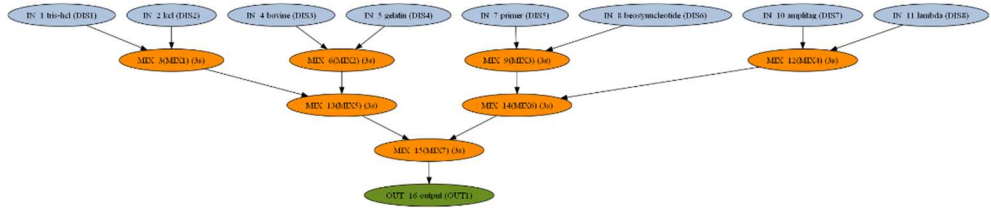

Figure 5. PCR Benchmark

- The in-vitro test refers to a series of diagnostic tests that use four human physiological substances or samples including plasma, serum, urine and saliva to determine the amount of glucose, lactate, pyruvate and glutamate to detect metabolic disorders [1]. In in-vitro testing, four physiological fluids are combined with four reactants, and the results are sent to a detection module embedded in the digital microfluidic biochip.

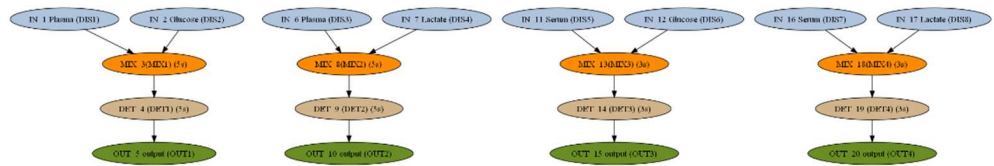

Figure 6. In-Vitro Benchmark

- Protein testing involves consecutive steps of mixing, splitting and detection. Depending on the number of bioassay nodes, we use seven different configurations of protein bioassays.

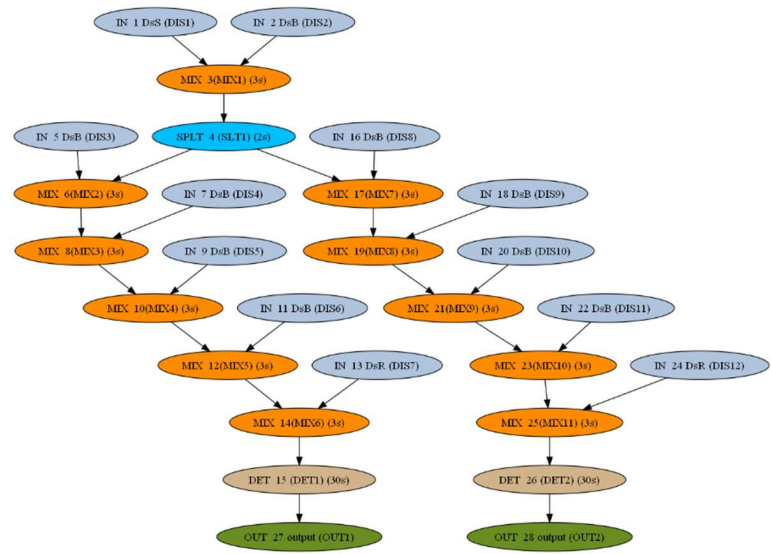

Figure 7. Protein Benchmark

Following the introduction of the benchmark bioassays for digital microfluidic biochips, we present the specification of the benchmark tests based on the number of operation (including mixing, splitting, storage, detection, etc.) in the form of a table.

Table 1. Benchmark Characteristics

| Benchmark | # Operations | | | | | | |
|---|---|---|---|---|---|---|---|
| | Dispense | Output | Mixing | Splitting | Detecting | Storage | Total |
| PCR | 8 | 1 | 7 | 0 | 0 | 0 | 16 |
| In-Vitro 1 | 8 | 4 | 4 | 0 | 4 | 0 | 20 |
| In-Vitro 2 | 12 | 6 | 6 | 0 | 6 | 0 | 30 |
| In-Vitro 3 | 18 | 9 | 9 | 0 | 9 | 0 | 45 |
| In-Vitro 4 | 24 | 12 | 12 | 0 | 12 | 0 | 60 |
| In-Vitro 5 | 32 | 16 | 16 | 0 | 16 | 0 | 80 |
| Protein 1 | 12 | 2 | 11 | 1 | 2 | 8 | 36 |
| Protein 2 | 24 | 4 | 23 | 3 | 4 | 18 | 76 |
| Protein 3 | 48 | 8 | 47 | 7 | 8 | 38 | 156 |
| Protein 4 | 96 | 16 | 95 | 15 | 16 | 78 | 316 |
| Protein 5 | 192 | 32 | 191 | 31 | 32 | 158 | 636 |
| Protein 6 | 384 | 64 | 383 | 63 | 64 | 318 | 1276 |
| Protein 7 | 768 | 128 | 767 | 127 | 128 | 638 | 2556 |

As shown in the table, the PCR test consists solely of mixing operations. The set of In-Vitro bioassays is composed of mixing and splitting operations; with the change in configuration from In-Vitro 1 to In-Vitro 5, the number of mixing and splitting operations increases proportionally. In relation to the set of protein bioassays, the mixing, splitting and storage operations are used; as shown in the storage column values, the protein set requires a large number of splitting/storage/detection modules to store the middle droplets.

# BENCHMARK ARCHITECTURE

The field-programmable pin-constrained (FPPC) architecture [10] is used as the benchmark architecture. The architecture is general-purpose and programmable, so it offers the capability to run a wide range of tests including PCR, In-Vitro and Protein.

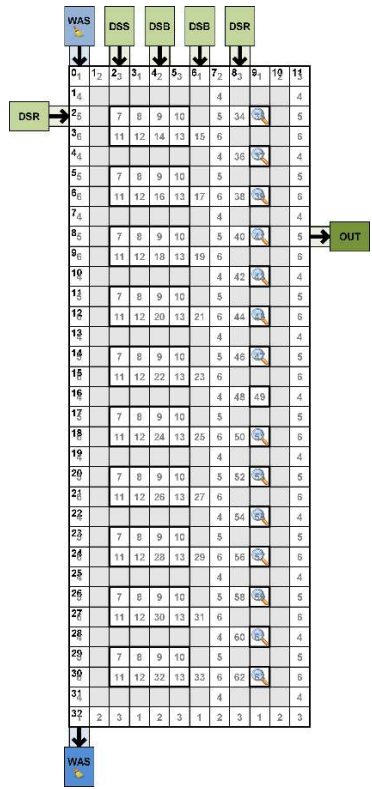

Figure 8. The FPPC Architecture

PROPOSED ARCHITECTURES

**The General-purpose Field-programmable Pin-Constrained (GFPC) Architecture**

The purpose of this architecture is to provide with minimum design size, number of electrodes and control pins, while also improving the performance level of digital microfluidic biochip. This has been accomplished by incorporating multiple paths for the routing of droplets on the surface of the digital microfluidic biochip [4].

Figure 9. The proposed GFPC Architecture

Before considering the performance of general-purpose field-programmable pin-constrained (GFPC) architecture, let's first consider the pin-mapping algorithm for the architecture. The pin-mapping pseudo-algorithm for the general-purpose field-programmable pin-constrained architecture is illustrated below.

The location of the vertical and horizontal routing columns should be determined according to the architectural dimensions; for this purpose, in lines 3 and 4, the variables *vrt* and *hrt* are defined to store the location of the vertical and horizontal routing columns.

```
Algorithm 1 Pin-mapping algorithm for GFPC architecture
 1: procedure GFPC_PM
 2:     /* Given the dimensions of the array Dim(x, y) */
 3:     Vector <int> vrt                                          ▷ Vertical routing columns
 4:     Vector <int> hrt                                          ▷ Horizontal routing columns
 5:
 6:     for ( ∀ i ∈ vrt ) do
 7:         Assign a pin number between 1 - 3;
 8:     end for
 9:
10:     for ( ∀ i ∈ hrt ) do
11:         Assign a pin number between 4 - 6;
12:     end for
13:
14:     function ALLOCATE_MIXING_RESOURCES( )
15:         Assign shared pins between 7 - 12 for pins inside the mixing modules;
16:         Assign an independent pin pin for I/O pin of the mixing module;
17:         Assign an independent pin pin for Hold pin of the mixing module;
18:     end function
19:
20:     function ALLOCATE_SSD_RESOURCES( )
21:         Assign an independent pin to SSD I/O pin;
22:         Assign an independent pin to SSD Hold pin;
23:     end function
24:
25:     function ALLOCATE_RB_RESOURCE( )
26:         Assign an independent pin to RB I/O pin;
27:         Assign an independent pin to RB Hold pin;
28:     end function
29: end procedure
```

After storing the location of the routing columns in the *vrt* and *hrt* variables, it is necessary to assign pins to each of the columns according to the type of column (vertical or horizontal).

In lines 6 to 9, each of the values in the *vrt* variable is read and assigned shared pins 1 to 3. In lines 10 through 12, each of the values stored in the *hrt* variable is read and assigned shared pins 4 to 6. Shared pins are used to reduce the total number of pins needed to address the electrodes of routing columns.

After assigning pins to the electrodes of the routing columns, we proceed to assign pins to the mixing modules. In lines 14 to 18, the electrodes within the mixing modules are assigned with the shared pins 7 to 12. Due to the need for separate controls for the I/O electrodes and the holder electrodes in the mixing modules, we assign separate pins to each I/O electrode and the holder.

In the following, pin assignment to splitting/storage/detection modules is discussed. For this purpose, lines 20 to 23 allocate independent pins to input/output electrodes and holders of each splitting/storage/detection modules.

Due to potential deadlocks in the routing process, a module is assigned as the routing buffer. In the last step of the pin mapping algorithm, lines 25 to 28 allocate independent pins to the I/O electrodes and holder for the routing buffer module.

In the following, we examine the performance of general-purpose field-programmable pin-constrained (GFPC) architecture in relation to the performance of the benchmark bioassays.

The benchmark test column shows the name of the benchmark bioassay applied to the architecture. The chip dimensions column represents the dimensions of the of the biochip electrodes for executing microfluidic operations.

The number of electrodes column represents the number of electrodes embedded in the surface of the digital microfluidic biochip. The pin number column indicates the number of control pins needed to control operations of the biochip.

The microfluidic operation time column shows the time allotted to perform digital microfluidic operations (such as merging, mixing, splitting, storage, detection, etc.).

The droplet routing column represents the droplet routing time on the surface of the electrodes during the bioassay execution. Droplet routing time includes droplet routing time from input dispensers to the modules, between the modules and from the modules to the output repositories.

Table 2. Comparing performance of the proposed GFPC architecture versus the FPPC benchmark architecture

| Bioassay | Dimensions | | # Electrodes | | # Pins | | Operations Time | | Routing Times | | Total Time | |
|---|---|---|---|---|---|---|---|---|---|---|---|---|
| | FP | GF | FP | GF | FP | GF | FP | GF | FP | GF | FP | GF |
| PCR | 21×12 | 11×16 | 153 | 126 | 43 | 42 | 11 | 11 | 2.1 | 1.6 | 13.1 | 12.6 |
| In-Vitro 1 | 21×12 | 11×16 | 153 | 126 | 43 | 42 | 14 | 14 | 2.6 | 2.1 | 16.6 | 16.1 |
| In-Vitro 2 | 21×12 | 11×16 | 153 | 126 | 43 | 42 | 18 | 18 | 3.8 | 3.0 | 21.8 | 21.0 |
| In-Vitro 3 | 21×12 | 11×16 | 153 | 126 | 43 | 42 | 18 | 18 | 6.2 | 4.7 | 24.2 | 22.7 |
| In-Vitro 4 | 21×12 | 11×16 | 153 | 126 | 43 | 42 | 19 | 19 | 8.8 | 6.5 | 27.8 | 25.5 |
| In-Vitro 5 | 21×12 | 11×16 | 153 | 126 | 43 | 42 | 25 | 25 | 11.6 | 8.4 | 36.6 | 33.4 |
| Protein 1 | 21×12 | 11×16 | 153 | 126 | 43 | 42 | 71 | 71 | 2.9 | 2.5 | 73.9 | 73.5 |
| Protein 2 | 21×12 | 11×16 | 153 | 126 | 43 | 42 | 106 | 106 | 6.1 | 5.2 | 112.1 | 111.2 |
| Protein 3 | 21×12 | 11×16 | 153 | 126 | 43 | 42 | 176 | 176 | 13.5 | 11.2 | 189.5 | 188.2 |
| Protein 4 | 21×12 | 11×16 | 153 | 126 | 43 | 42 | 316 | 316 | 29.3 | 24.4 | 345.3 | 340.4 |
| Protein 5 | 25×12 | 11×16 | 178 | 130 | 49 | 46 | 596 | 596 | 61.4 | 51.6 | 657.4 | 647.6 |
| Protein 6 | 29×12 | 11×21 | 203 | 165 | 55 | 54 | 1156 | 1156 | 127.4 | 107.0 | 1283.4 | 1263.0 |
| Protein 7 | 33×12 | 11×21 | 237 | 169 | 63 | 58 | 2276 | 2276 | 260.6 | 217.8 | 2536.6 | 2493.8 |
| Improvement | 1.33 | | 1.20 | | 1.03 | | 1.00 | | 1.17 | | 1.02 | |

The total time column represents the total time the bioassay was performed on digital microfluidic biochip. The total time is obtained from the sum of the columns of the operation time and the droplet routing time.

With regard to the dimensions of digital microfluidic biochip, the results show a 33% improvement (which means that the dimensions are reduced). The results also show a 20% decrease in the number of electrodes. Results show a 3% decrease in the number of control pins.

Both architectures provide the same performance, while droplet routing is down 17%. The table shows that the droplet routing times are much shorter compared to the time of the microfluidic operations, although due to the equality of the time of the microfluidic operations in both architectures, the reduction in droplet routing time is significant and reduces the total time required for bioassay execution by 2%.

**Field-programmable Pin-constrained Cell Array (FPCA) Architecture**

In the design of field-programmable pin-constrained Cell Array (FPCA) [5] architecture the cells are utilized as the basic structure; each cell comprises of a certain number of mixing modules as well as splitting/storage/detection modules. The overall structure of the FPCA architecture is a constructed of number of cells. For example, the FPCA architecture presented below is made up of four basic cells. Each cell has two mixing modules and four splitting/storage/detection modules.

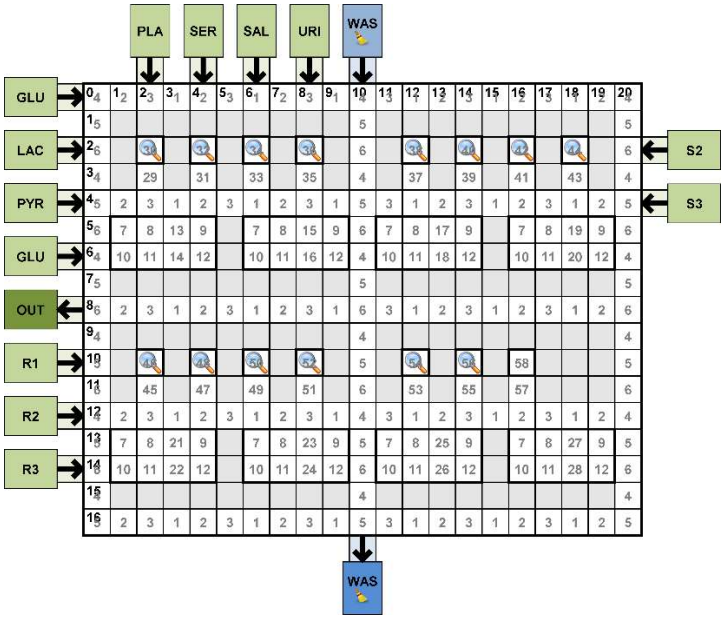

Figure 10. The proposed FPCA Architecture

Before examining the performance of FPCA architecture, we first examine the pin-mapping algorithm for the architecture in question. The pin-mapping pseudo-algorithm is shown below.

```
Algorithm 2 Pin-mapping algorithm for FPCA architecture
 1: procedure FPCA_PM
 2:     /* Given the dimensions of the array Dim(x, y) */
 3:     Vector <int> vrt                                              ▷ Vertical routing columns
 4:     Vector <int> hrt                                              ▷ Horizontal routing columns
 5:
 6:     for ( ∀ i ∈ vrt ) do
 7:         Assign a pin number between 1 - 3;
 8:     end for
 9:
10:     for ( ∀ i ∈ hrt ) do
11:         Assign a pin number between 4 - 6;
12:     end for
13:
14:     function ALLOCATE_MIXING_RESOURCES( )
15:         Assign shared pins between 7 - 12 for pins inside the mixing modules;
16:         Assign an independent pin pin for I/O pin of the mixing module;
17:         Assign an independent pin pin for Hold pin of the mixing module;
18:     end function
19:
20:     function ALLOCATE_SSD_RESOURCES( )
21:         Assign an independent pin to SSD I/O pin;
22:         Assign an independent pin to SSD Hold pin;
23:     end function
24:
25:     function ALLOCATE_RB_RESOURCE( )
26:         Assign an independent pin to RB I/O pin;
27:         Assign an independent pin to RB Hold pin;
28:     end function
29: end procedure
```

The location of the vertical and horizontal routing columns should be determined according to the architectural dimensions; for this purpose, in lines 3 and 4, the variables *vrt* and *hrt* are defined to store the location of the vertical and horizontal routing columns.

After storing the location of the routing columns in the *vrt* and *hrt* variables, it is necessary to assign pins to each of the columns according to the type of column (vertical or horizontal).

In lines 6 to 9, each of the values in the *vrt* variable is first read and then assigned to the shared pins 1 to 3. Shared pins are used to reduce the total number of pins needed to address the electrodes of routing columns.

In lines 10 through 12, each of the values stored in the *hrt* variable is read and then assigned to the 4 to 6 shared pins. Shared pins are used to reduce the total number of pins needed to address the electrodes of routing columns.

In the following, we investigate the performance of the FPCA architecture in relation to the performance of the benchmark FPPC architecture.

The benchmark test column shows the name of the benchmark bioassay applied to the architecture. The chip dimensions column represents the dimensions of the of the biochip electrodes for executing microfluidic operations.

The number of electrodes column represents the number of electrodes embedded in the surface of the digital microfluidic biochip. The pin number column indicates the number of control pins needed to control operations of the biochip.

The microfluidic operation time column shows the time allotted to perform digital microfluidic operations (such as merging, mixing, splitting, storage, detection, etc.).

The droplet routing column represents the droplet routing time on the surface of the electrodes during the bioassay execution. Droplet routing time includes droplet routing time from input reservoirs to the modules, between the modules and from the modules to the output reservoirs.

The total time column represents the total time the bioassay was performed on digital microfluidic biochip. The total time is obtained from the sum of the columns of the operation time and the droplet routing time.

Here's a look at how different parameters of the FPCA architecture are compared to those of the FPPC architecture.

Table 3. Comparing performance of the proposed FPCA architecture versus the FPPC benchmark architecture

| Bioassay | Dimensions | | # Electrodes | | # Pins | | Operations Time | | Routing Times | | Total Time | |
|---|---|---|---|---|---|---|---|---|---|---|---|---|
| | FP | CA | FP | CA | FP | CA | FP | CA | FP | CA | FP | CA |
| PCR | 33×12 | 17×21 | 237 | 235 | 63 | 58 | 11 | 11 | 2.1 | 1.8 | 13.1 | 12.8 |
| In-Vitro 1 | 33×12 | 17×21 | 237 | 235 | 63 | 58 | 14 | 14 | 2.6 | 2.2 | 16.6 | 16.2 |
| In-Vitro 2 | 33×12 | 17×21 | 237 | 235 | 63 | 58 | 18 | 18 | 3.8 | 3.0 | 21.8 | 21.0 |
| In-Vitro 3 | 33×12 | 17×21 | 237 | 235 | 63 | 58 | 18 | 18 | 6.2 | 5.1 | 24.2 | 23.1 |
| In-Vitro 4 | 33×12 | 17×21 | 237 | 235 | 63 | 58 | 18 | 18 | 9.4 | 7.5 | 27.4 | 25.5 |
| In-Vitro 5 | 33×12 | 17×21 | 237 | 235 | 63 | 58 | 20 | 21 | 14.5 | 10.7 | 34.5 | 31.7 |
| Protein 1 | 33×12 | 17×21 | 237 | 235 | 63 | 58 | 71 | 71 | 2.9 | 2.5 | 73.9 | 73.5 |
| Protein 2 | 33×12 | 17×21 | 237 | 235 | 63 | 58 | 106 | 106 | 6.1 | 5.2 | 112.1 | 111.2 |
| Protein 3 | 33×12 | 17×21 | 237 | 235 | 63 | 58 | 176 | 176 | 13.2 | 11.6 | 189.2 | 187.6 |
| Protein 4 | 33×12 | 17×21 | 237 | 235 | 63 | 58 | 316 | 316 | 28.4 | 24.9 | 344.4 | 340.9 |
| Protein 5 | 33×12 | 17×21 | 237 | 235 | 63 | 58 | 596 | 596 | 60.4 | 53.6 | 656.4 | 649.6 |
| Protein 6 | 33×12 | 17×21 | 237 | 235 | 63 | 58 | 1156 | 1156 | 126.5 | 112.4 | 1282.5 | 1268.4 |
| Protein 7 | 33×12 | 17×21 | 237 | 235 | 63 | 58 | 2276 | 2276 | 260.6 | 230.2 | 2536.6 | 2506.2 |
| Improvement | 1.10 | | 1.01 | | 1.08 | | 1.00 | | 1.12 | | 1.01 | |

With regard to the dimensions of digital microfluidic biochip dimensions the results show a 10% improvement. It also shows a 1% decrease in the number of electrodes. Results show an 8% decrease in the number of control pins.

Both architectures provide the same performance when it comes to microfluidic operations. There is a 12% reduction in droplet routing time. According to the table, it is observed that droplet routing times are much shorter compared to the time of microfluidic operations, although due to the equality of time of microfluidic operations in both architectures, the reduction in droplet routing time in the FPCA architecture is considerable compared to the FPPC architecture and reduces the total bioassay time by 1%.

## FAULT-TOLERANCE IN PROPOSED ARCHITECTURES

In this section, we investigate the fault-tolerance in the presented GFPC and FPCA architectures and compare the fault-tolerance performance in the aforementioned architectures with respect to the fault-tolerance in the FPPC architecture.

The fault-tolerance capability of in GFPC [4] and FPCA [5] architectures includes fault-tolerance in mixing modules, splitting/storage/detection modules, and droplet routing paths; as we will discuss each of these in detail later.

**Fault-Tolerance in Mixing Modules**

In this section, we describe the failure of mixing module and its impact on the bioassay process.

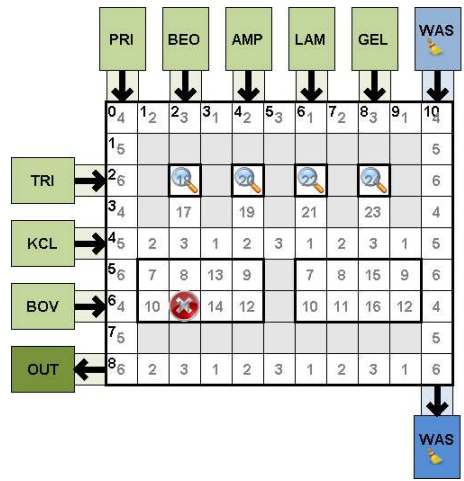

Figure 11. Faults affecting mixing modules

The use of mixing modules affected by permanent faults can disrupt the process of performing the bioassay. Simply put, if the electrode in the mixing module is defective when the droplet passes through the defective electrode, it will not be possible to activate and apply the electric field to that electrode and thus the droplet will not be able to move.

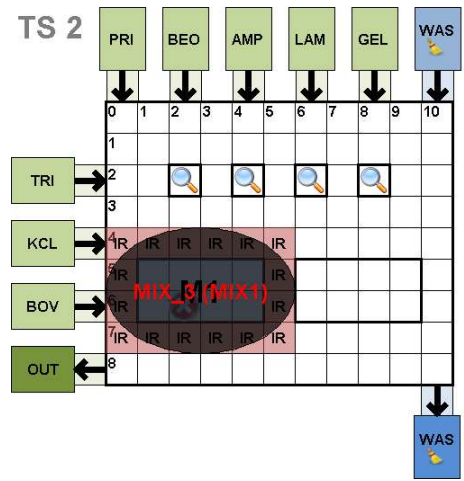

Figure 12. Placement of mixing microfluidic operation on faulty mixing modules

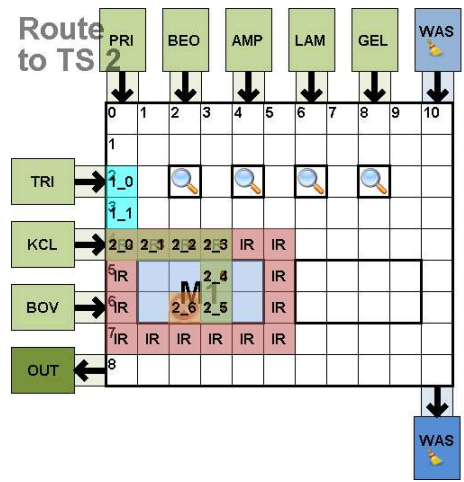

Figure 13. Droplet routing of mixing microfluidic operation on faulty mixing modules

Therefore, the processing of bioassay is interrupted. If the permanent fault in the mixing module electrode is not detected prior to performing the bioassay synthesis process, the faulty mixing module will be used in the bioassay execution process and thus the execution process will be disrupted. The proposed solution is to modify the existing algorithms in order to skip the defective mixing module from the set of modules for bioassay execution.

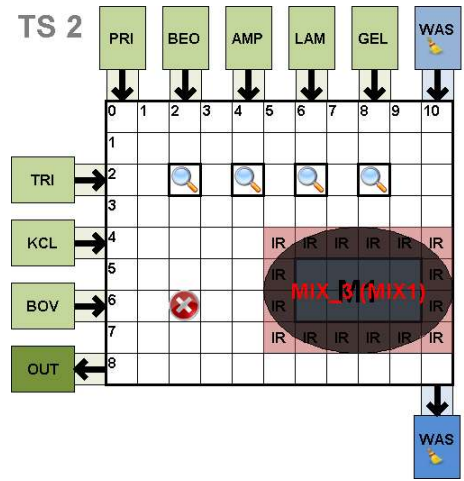

Figure 14. Placement of mixing microfluidic operation skipping faulty mixing modules

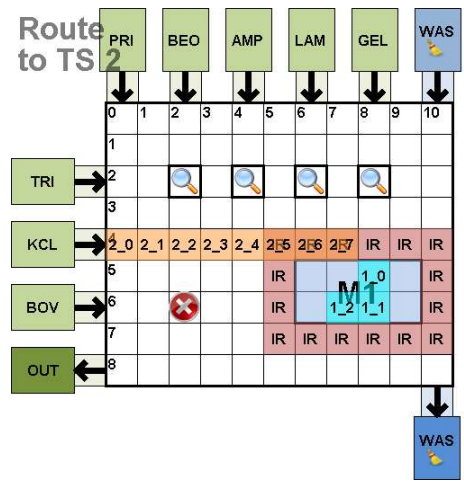

Figure 15. Droplet routing of mixing microfluidic operation skipping faulty mixing modules

**Fault-Tolerance in Splitting/Storage/Detection Modules**

In this section we describe the permanent faults in the division/storage/detection module and its impact on the bioassay execution process. If there are any permanent faults in the splitting/storage/detection module electrode prior to performing the bioassay synthesis process, it will result in disruption of the test execution process.

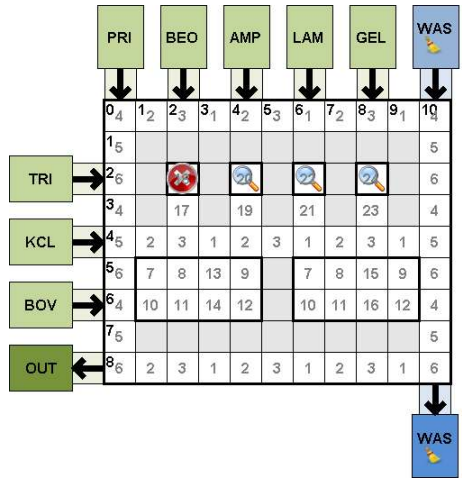

Figure 16. Faults affecting splitting/storage/detection modules

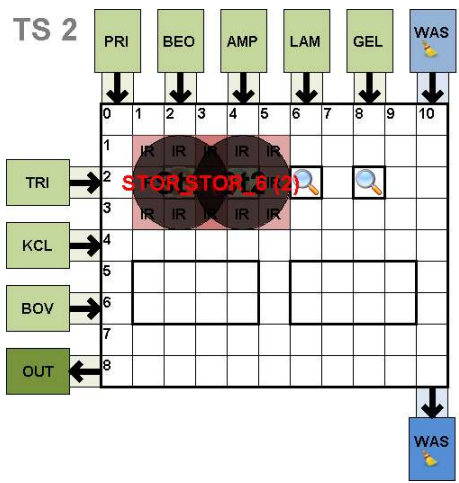

Figure 17. Placement of microfluidic operation on faulty splitting/storage/detection modules

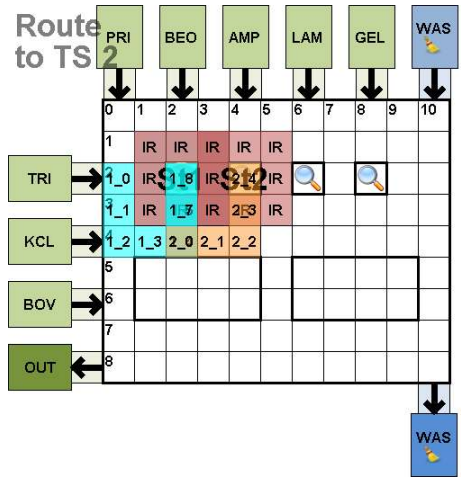

Figure 18. Droplet routing of microfluidic operation on faulty splitting/storage/detection modules

The proposed solution is to modify the existing algorithms in such a way as to skip the faulty splitting/storage/detection modules from the set of modules available for bioassay execution; therefore, no operations are inserted and thus there will be no droplets in the faulty splitting/storage/detection module.

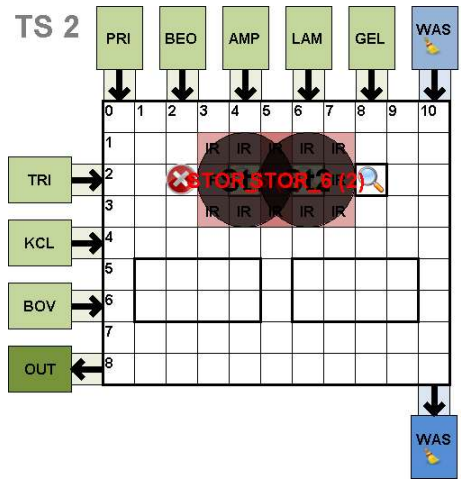

Figure 19. Placement of microfluidic operation skipping faulty splitting/storage/detection modules

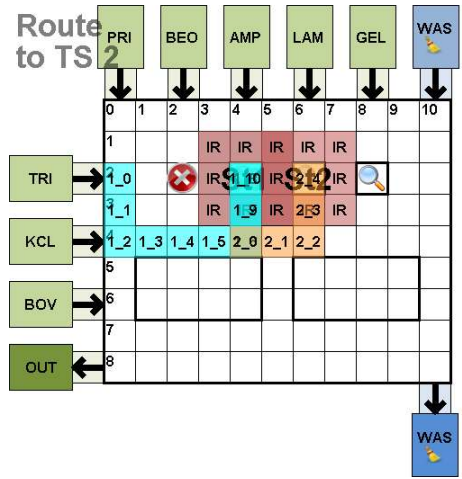

Figure 20. Droplet routing of microfluidic operation skipping faulty splitting/storage/detection modules

**Fault-Tolerance of Droplet Routing Pathways**

Regarding the fault-tolerance of routing paths, since the proposed architectures provide multiple paths for droplet routing, alternative embedded paths can be used in the event of a permanent breakdown at one of the routing pathway electrodes. Whereas in the benchmark architecture there is only one path for droplet routing and any breakdown in the pathway electrodes will affect the performance of the DMFB architecture.

**Conclusion**

In this chapter, we first introduced a set of benchmark tests in relation to DMFB devices, and then presented the proposed fault-tolerant architectures. Then we compared the performance of the proposed architectures with the benchmark FPPC architecture. Finally, we explained the concepts of fault-tolerance in different parts of DMFB architectures and how to deal with the occurrence of permanent faults in each segment.

As detailed comparisons of the various parameters of the proposed architectures showed, the proposed architectures provide a better performance along with improved fault-tolerance.

*Chapter 3*

FAULT-TOLERANCE EVALUATION RESULTS

After introducing the proposed architectures in the previous chapter, this chapter examines the performance of the proposed architectures in the event of a breakdown in different parts of the DMFB architecture. We first compare the performance of the proposed architectures in the presence of the failure of the mixing module. Next, we examine the performance of the proposed architectures compared to the benchmark architecture in the presence of splitting/storage/detection module failures. Finally, we refer to fault-tolerance in routing buses [6].

As can be seen in the following table, in the case of the GFPC and FPCA architectures, the dimensions of the array are 21 × 11 and 21 × 17, with respectively, 169 and 235 electrodes and 237 and 58 control pins.

As shown in table, in case of GFPC architecture, there is a 42% reduction in array dimensions, 29% decrease in the number of electrodes, and an 8% decrease in the number of pins compared to the FPPC architecture. In case of FPCA architecture there is a 10% reduction in array size, 1% decrease in the number of electrodes and an 8% decrease in the number of control pins.

Table 4. Characteristics of DMFB architectures; GFPC versus FPPC versus FPCA

| Benchmarks | Dimensions | | | # Electrodes | | | # Pins | | |
|---|---|---|---|---|---|---|---|---|---|
| | GF | FP | CA | GF | GP | CA | GF | FP | CA |
| PCR | 11×21 | 33×12 | 17×21 | 169 | 237 | 235 | 58 | 63 | 58 |
| In-Vitro 1 | 11×21 | 33×12 | 17×21 | 169 | 237 | 235 | 58 | 63 | 58 |
| In-Vitro 2 | 11×21 | 33×12 | 17×21 | 169 | 237 | 235 | 58 | 63 | 58 |
| In-Vitro 3 | 11×21 | 33×12 | 17×21 | 169 | 237 | 235 | 58 | 63 | 58 |
| In-Vitro 4 | 11×21 | 33×12 | 17×21 | 169 | 237 | 235 | 58 | 63 | 58 |
| In-Vitro 5 | 11×21 | 33×12 | 17×21 | 169 | 237 | 235 | 58 | 63 | 58 |
| Protein 1 | 11×21 | 33×12 | 17×21 | 169 | 237 | 235 | 58 | 63 | 58 |
| Protein 2 | 11×21 | 33×12 | 17×21 | 169 | 237 | 235 | 58 | 63 | 58 |
| Protein 3 | 11×21 | 33×12 | 17×21 | 169 | 237 | 235 | 58 | 63 | 58 |
| Protein 4 | 11×21 | 33×12 | 17×21 | 169 | 237 | 235 | 58 | 63 | 58 |
| Protein 5 | 11×21 | 33×12 | 17×21 | 169 | 237 | 235 | 58 | 63 | 58 |
| Protein 6 | 11×21 | 33×12 | 17×21 | 169 | 237 | 235 | 58 | 63 | 58 |
| Protein 7 | 11×21 | 33×12 | 17×21 | 169 | 237 | 235 | 58 | 63 | 58 |
| Improvement | 1.42 / 1.10 | | | 1.29 / 1.01 | | | 1.08 / 1.08 | | |

In simulating the occurrence of permanent faults in the proposed architectures, only the first (most used) module of any type (mixing module, or splitting/storage/detection module) is assumed to be faulty.

**Failure in Mixing Modules**

In DMFB architectures, a group of shared pins are assigned to the modules; therefore, even if one of the mixing modules fails, it may be time-consuming or even impossible to use. Therefore, we modify the existing algorithms so that the synthesis flow skips the faulty module, and instead uses other existing modules if possible.

**Comparison of Mixing Module Failures in GFPC Architecture versus the FPPC Benchmark Architecture**

Previously, we compared the performance of the proposed GFPC architecture and the FPPC benchmark architecture in the normal conditions without the occurrence of permanent faults. In this section, we intend to examine the performance of the aforementioned architectures in the presence of permanent faults in the mixing module [6]. It should be noted that in connection with the assumption of permanent malfunction of the mixing module, we used the first module (the most frequently used module) as the faulty module.

The benchmark column represents the type of bioassay applied on the DMFB. The operation time column represents the time allocated to performing digital microfluidic operations, including mixing, splitting, storage, detection, etc. Each of the healthy and faulty sub-columns are further divided into FP and GF. Healthy represents the normal operation time and no faults. In contrast, the Faulty sub-column represents the operation time in the event of a permanent fault in the mixing module in the GFPC and FPPC architectures.

The routing time column represents the time allotted for routing the droplets from the input reservoirs to the modules, between the modules and from the modules to the output reservoirs. The routing column is divided into two healthy and faulty sub-columns; each of the healthy and faulty sub-columns is in turn divided into two FP and GF. The healthy sub-column represents the droplet routing time in the normal state without any faults. Faulty sub-column indicates the droplet routing time in the event of a permanent fault in the mixing module in the GFPC (GF) and benchmark FPPC (FP) architectures.

The total time column represents the total time of the test and, in other words, the sum of the columns of the operation time and droplet routing time.

As shown in the table the two bottom rows indicate the average improvement and the average overhead. The average overhead row represents the overhead rate compared to the performance of each of the architectures in both normal and faulty performance modes.

Table 5. Comparing performance of GFPC versus FPPC architectures in presence of faulty mixing module

| Benchmark | Operation time | | | | Droplet Routing Time | | | | Total Time | | | |
|---|---|---|---|---|---|---|---|---|---|---|---|---|
| | Healthy | | Faulty | | Healthy | | Faulty | | Healthy | | Faulty | |
| | FP | GF | FP | GF | FP | GF | FP | GF | FP | GF | FP | GF |
| PCR | 11 | 11 | 11 | 11 | 2.1 | 1.9 | 2.4 | 1.8 | 13.1 | 12.9 | 13.4 | 12.8 |
| In-Vitro 1 | 14 | 14 | 14 | 14 | 2.6 | 2.2 | 2.9 | 2.3 | 16.6 | 16.2 | 16.9 | 16.3 |
| In-Vitro 2 | 18 | 18 | 18 | 18 | 3.8 | 3.1 | 4.3 | 3.3 | 21.8 | 21.1 | 22.3 | 21.3 |
| In-Vitro 3 | 18 | 18 | 18 | 18 | 6.2 | 5.3 | 6.8 | 5.2 | 24.2 | 23.3 | 24.8 | 23.2 |
| In-Vitro 4 | 18 | 18 | 18 | 18 | 9.4 | 6.6 | 10.3 | 7.1 | 27.4 | 24.6 | 28.3 | 25.1 |
| In-Vitro 5 | 20 | 21 | 21 | 23 | 14.5 | 10 | 14.6 | 10.2 | 34.5 | 31.0 | 35.6 | 33.2 |
| Protein 1 | 71 | 71 | 71 | 71 | 2.9 | 2.5 | 3.7 | 3.9 | 73.9 | 73.5 | 74.7 | 74.9 |
| Protein 2 | 106 | 106 | 106 | 106 | 6.1 | 5.3 | 7.2 | 7.6 | 112.1 | 111.3 | 113.2 | 113.6 |
| Protein 3 | 176 | 176 | 176 | 176 | 13.2 | 11.4 | 14.9 | 15 | 189.2 | 187.4 | 190.9 | 191.0 |
| Protein 4 | 316 | 316 | 316 | 316 | 28.4 | 24.0 | 30.7 | 30.0 | 344.4 | 340.0 | 346.7 | 346.0 |
| Protein 5 | 596 | 596 | 596 | 596 | 60.4 | 51.6 | 63.9 | 62.4 | 656.4 | 647.6 | 659.9 | 658.4 |
| Protein 6 | 1156 | 1156 | 1156 | 1156 | 126.5 | 106.2 | 132.3 | 127.3 | 1282.5 | 1262.2 | 1288.3 | 1283.3 |
| Protein 7 | 2276 | 2276 | 2276 | 2276 | 260.6 | 215.7 | 271.5 | 259.3 | 2536.6 | 2491.7 | 2547.5 | 2535.3 |
| Improvement | 1.00 | | 1.00 | | 1.17 | | 1.05 | | 1.02 | | 1.01 | |
| Overhead | 1.00 = FP | | | | 1.05 = FP | | | | 1.01 = FP | | | |
| | 1.00 = GF | | | | 1.20 = GF | | | | 1.02 = GF | | | |

**Comparison of Mixing Module Failures in FPCA Architecture versus the FPPC Benchmark Architecture**

In this section, we intend to examine the performance of the FPCA and benchmark FPPC architectures in the presence of permanent faults in the mixing module [6]. It should be noted that in connection with the assumption of permanent malfunction of the mixing module, we used the first module (the most frequently used module) as the faulty module.

The benchmark column represents the type of bioassay applied on the DMFB. The operation time column represents the time allocated to performing digital microfluidic operations, including mixing, splitting, storage, detection, etc. Each of the healthy and defective sub-columns are further divided into FP and CA. Healthy represents the normal operation time and no faults. In contrast, the Faulty sub-column represents the operation time in the event of a permanent fault in the mixing module in the FPCA and FPPC architectures.

The routing time column represents the time allotted for routing the droplets from the input reservoirs to the modules, between the modules and from the modules to the output reservoirs. The routing column is divided into two healthy and faulty sub-columns; each of the healthy and faulty sub-columns is in turn divided into two FP and CA. The healthy sub-column represents the droplet routing time in the normal state without any faults. Faulty sub-column indicates the droplet routing time in the event of a permanent fault in the mixing module in the FPCA (CA) and benchmark FPPC (FP) architectures.

The total time column represents the total time of the test and, in other words, the sum of the columns of the operation time and droplet routing time.

As shown in the table the two bottom rows indicate the average improvement and the average overhead. The average overhead row represents the overhead rate compared to the performance of each of the architectures in both normal and faulty performance modes.

Table 6. Comparing performance of FPCA versus FPPC architectures in presence of faulty mixing module

| Benchmark | Operation time | | | | Droplet Routing Time | | | | Total Time | | | |
|---|---|---|---|---|---|---|---|---|---|---|---|---|
| | Healthy | | Faulty | | Healthy | | Faulty | | Healthy | | Faulty | |
| | FP | CA | FP | CA | FP | CA | FP | CA | FP | CA | FP | CA |
| PCR | 11 | 11 | 11 | 11 | 2.1 | 1.8 | 2.4 | 2.2 | 13.1 | 12.8 | 13.4 | 13.2 |
| In-Vitro 1 | 14 | 14 | 14 | 14 | 2.6 | 2.2 | 2.9 | 2.6 | 16.6 | 16.2 | 16.9 | 16.6 |
| In-Vitro 2 | 18 | 18 | 18 | 18 | 3.8 | 3.0 | 4.3 | 3.7 | 21.8 | 21.0 | 22.3 | 21.7 |
| In-Vitro 3 | 18 | 18 | 18 | 18 | 6.2 | 5.1 | 6.8 | 5.8 | 24.2 | 23.1 | 24.8 | 23.8 |
| In-Vitro 4 | 18 | 18 | 18 | 18 | 9.4 | 7.5 | 10.3 | 8.5 | 27.4 | 25.5 | 28.3 | 26.5 |
| In-Vitro 5 | 20 | 21 | 21 | 23 | 14.5 | 10.7 | 14.6 | 11.8 | 34.5 | 31.7 | 35.6 | 34.8 |
| Protein 1 | 71 | 71 | 71 | 71 | 2.9 | 2.5 | 3.7 | 3.9 | 73.9 | 73.5 | 74.7 | 74.9 |
| Protein 2 | 106 | 106 | 106 | 106 | 6.1 | 5.3 | 7.2 | 7.6 | 112.1 | 111.2 | 113.2 | 113.6 |
| Protein 3 | 176 | 176 | 176 | 176 | 13.2 | 11.4 | 14.9 | 15.1 | 189.2 | 187.6 | 190.9 | 191.1 |
| Protein 4 | 316 | 316 | 316 | 316 | 28.4 | 24.0 | 30.7 | 30.1 | 344.4 | 340.9 | 346.7 | 346.1 |
| Protein 5 | 596 | 596 | 596 | 596 | 60.4 | 51.6 | 63.9 | 63.1 | 656.4 | 649.6 | 659.9 | 659.1 |
| Protein 6 | 1156 | 1156 | 1156 | 1156 | 126.5 | 106.2 | 132.3 | 130.8 | 1282.5 | 1268.4 | 1288.3 | 1286.8 |
| Protein 7 | 2276 | 2276 | 2276 | 2276 | 260.6 | 215.7 | 271.5 | 266.8 | 2536.6 | 2506.2 | 2547.5 | 2542.8 |
| Improvement | 1.00 | | 1.00 | | 1.12 | | 1.02 | | 1.01 | | 1.00 | |
| Overhead | 1.00 = FP | | | | 1.05 = FP | | | | 1.01 = FP | | | |
| | 1.00 = CA | | | | 1.15 = CA | | | | 1.02 = CA | | | |

**Failure in Splitting/Storage/Detection Modules**

In this section, we discuss the splitting/storage/detection module failure and its impact on the performance of proposed architectures. In the case of splitting/storing/detection modules if the input electrode or the holding electrode fails, the only way is not to use the module [6]. Next, we investigate the effect of splitting/storage/detection module failure on the performance of architectures.

**Comparison of Splitting/Storage/Detection Module Failures in GFPC Architecture versus the FPPC Benchmark Architecture**

In this section, we intend to examine the performance of the aforementioned architectures in the presence of permanent faults in the splitting/storage/detection module. It should be noted that in connection with the assumption of permanent malfunction of the splitting/storage/detection module, we used the first module (the most frequently used module) as the faulty module.

The benchmark column represents the type of bioassay applied on the DMFB. The operation time column represents the time allocated to performing digital microfluidic operations, including mixing, splitting, storage, detection, etc. Each of the healthy and faulty sub-columns are further divided into FP and GF. Healthy represents the normal operation time and no faults. In contrast, the Faulty sub-column represents the operation time in the event of a permanent fault in the splitting/storage/detection module in the GFPC and FPPC architectures.

The routing time column represents the time allotted for routing the droplets from the input reservoirs to the modules, between the modules and from the modules to the output reservoirs. The routing column is divided into two healthy and faulty sub-columns; each of the healthy and faulty sub-columns is in turn divided into two FP and GF. The healthy sub-column represents the droplet routing time in the normal state without any faults. Faulty sub-column indicates the droplet routing time in the event of a permanent fault in the splitting/storage/detection module in the GFPC (GF) and benchmark FPPC (FP) architectures.

Table 7. Comparing performance of GFPC versus FPPC architectures in presence of faulty splitting/storage/detection module

| Benchmark | Operation time | | | | Droplet Routing Time | | | | Total Time | | | |
|---|---|---|---|---|---|---|---|---|---|---|---|---|
| | Healthy | | Faulty | | Healthy | | Faulty | | Healthy | | Faulty | |
| | FP | GF | FP | GF | FP | GF | FP | GF | FP | GF | FP | GF |
| PCR | 11 | 11 | 11 | 11 | 2.1 | 1.9 | 2.1 | 1.9 | 13.1 | 12.9 | 13.1 | 12.9 |
| In-Vitro 1 | 14 | 14 | 14 | 14 | 2.6 | 2.2 | 2.6 | 2.2 | 16.6 | 16.2 | 16.6 | 16.2 |
| In-Vitro 2 | 18 | 18 | 18 | 18 | 3.8 | 3.1 | 4.0 | 3.3 | 21.8 | 21.1 | 22.0 | 21.3 |
| In-Vitro 3 | 18 | 18 | 18 | 18 | 6.2 | 5.3 | 6.3 | 4.8 | 24.2 | 23.3 | 24.3 | 22.8 |
| In-Vitro 4 | 18 | 18 | 18 | 18 | 9.4 | 6.6 | 9.7 | 6.8 | 27.4 | 24.6 | 27.7 | 24.8 |
| In-Vitro 5 | 20 | 21 | 20 | 21 | 14.5 | 10.0 | 14.8 | 9.3 | 34.5 | 31.0 | 34.8 | 30.3 |
| Protein 1 | 71 | 71 | 71 | 71 | 2.9 | 2.5 | 2.7 | 2.4 | 73.9 | 73.5 | 73.7 | 73.4 |
| Protein 2 | 106 | 106 | 106 | 106 | 6.1 | 5.3 | 6.3 | 5.4 | 112.1 | 111.3 | 112.3 | 111.4 |
| Protein 3 | 176 | 176 | 176 | 176 | 13.2 | 11.4 | 14.1 | 11.9 | 189.2 | 187.4 | 190.1 | 187.9 |
| Protein 4 | 316 | 316 | 316 | 316 | 28.4 | 24.0 | 30.9 | 26.0 | 344.4 | 340.0 | 346.9 | 342.0 |
| Protein 5 | 596 | 596 | 596 | 596 | 60.4 | 51.6 | 65.9 | 54.4 | 656.4 | 647.6 | 661.9 | 650.4 |
| Protein 6 | 1156 | 1156 | 1156 | 1156 | 126.5 | 106.2 | 137.7 | 111.7 | 1282.5 | 1262.2 | 1293.7 | 1267.7 |
| Protein 7 | *** | *** | *** | *** | *** | *** | *** | *** | *** | *** | *** | *** |
| Improvement | 1.00 | | 1.00 | | 1.17 | | 1.19 | | 1.02 | | 1.02 | |
| Overhead | 1.00 = FP | | | | 1.07 = FP | | | | 1.01 = FP | | | |
| | 1.00 = GF | | | | 1.04 = GF | | | | 1.00 = GF | | | |

The total time column represents the total time of the test and, in other words, the sum of the columns of the operation time and droplet routing time.

As shown in the table the two bottom rows indicate the average improvement and the average overhead. The average overhead row represents the overhead rate compared to the performance of each of the architectures in both normal and faulty performance modes.

**Comparison of Splitting/Storage/Detection Module Failures in FPCA Architecture versus the FPPC Benchmark Architecture**

In this section, we intend to examine the performance of the aforementioned architectures in the presence of permanent faults in the splitting/storage/detection module [6]. It should be noted that in connection with the assumption of permanent malfunction of the splitting/storage/detection module, we used the first module (the most frequently used module) as the faulty module.

The benchmark column represents the type of bioassay applied on the DMFB. The operation time column represents the time allocated to performing digital microfluidic operations, including mixing, splitting, storage, detection, etc. Each of the healthy and faulty sub-columns are further divided into FP and CA. Healthy represents the normal operation time and no faults. In contrast, the Faulty sub-column represents the operation time in the event of a permanent fault in the splitting/storage/detection module in the FPCA and FPPC architectures.

The routing time column represents the time allotted for routing the droplets from the input reservoirs to the modules, between the modules and from the modules to the output reservoirs. The routing column is divided into two healthy and faulty sub-columns; each of the healthy and faulty sub-columns is in turn divided into two FP and CA. The healthy sub-column represents the droplet routing time in the normal state without any faults. Faulty sub-column indicates the droplet routing time in the event of a permanent fault in the splitting/storage/detection module in the FPCA (CA) and benchmark FPPC (FP) architectures.

Table 8. Comparing performance of FPCA versus FPPC architectures in presence of faulty splitting/storage/detection module

| Benchmark | Operation time | | | | Droplet Routing Time | | | | Total Time | | | |
|---|---|---|---|---|---|---|---|---|---|---|---|---|
| | Healthy | | Faulty | | Healthy | | Faulty | | Healthy | | Faulty | |
| | FP | CA | FP | CA | FP | CA | FP | CA | FP | CA | FP | CA |
| PCR | 11 | 11 | 11 | 11 | 2.1 | 1.8 | 2.1 | 1.8 | 13.1 | 12.8 | 13.1 | 12.8 |
| In-Vitro 1 | 14 | 14 | 14 | 14 | 2.6 | 2.2 | 2.6 | 2.3 | 16.6 | 16.2 | 16.6 | 16.3 |
| In-Vitro 2 | 18 | 18 | 18 | 18 | 3.8 | 3.0 | 4.0 | 3.2 | 21.8 | 21.1 | 22.0 | 21.2 |
| In-Vitro 3 | 18 | 18 | 18 | 18 | 6.2 | 5.1 | 6.3 | 5.2 | 24.2 | 23.3 | 24.3 | 23.2 |
| In-Vitro 4 | 18 | 18 | 18 | 18 | 9.4 | 7.5 | 9.7 | 7.9 | 27.4 | 24.6 | 27.7 | 25.9 |
| In-Vitro 5 | 20 | 21 | 20 | 21 | 14.5 | 10.7 | 14.8 | 11.1 | 34.5 | 31.0 | 34.8 | 32.1 |
| Protein 1 | 71 | 71 | 71 | 71 | 2.9 | 2.5 | 2.7 | 2.4 | 73.9 | 73.5 | 73.7 | 73.4 |
| Protein 2 | 106 | 106 | 106 | 106 | 6.1 | 5.2 | 6.3 | 5.5 | 112.1 | 111.3 | 112.3 | 111.5 |
| Protein 3 | 176 | 176 | 176 | 176 | 13.2 | 11.6 | 14.1 | 12.4 | 189.2 | 187.4 | 190.1 | 188.4 |
| Protein 4 | 316 | 316 | 316 | 316 | 28.4 | 24.9 | 30.9 | 28.5 | 344.4 | 340.0 | 346.9 | 344.5 |
| Protein 5 | 596 | 596 | 596 | 596 | 60.4 | 53.6 | 65.9 | 60.6 | 656.4 | 647.6 | 661.9 | 656.6 |
| Protein 6 | 1156 | 1156 | 1156 | 1156 | 126.5 | 112.4 | 137.7 | 125.2 | 1282.5 | 1262.2 | 1293.7 | 1281.2 |
| Protein 7 | 2276 | 2276 | *** | *** | 260.6 | 230.2 | *** | *** | 2536.6 | 2506.2 | *** | *** |
| Improvement | 1.00 | | 1.00 | | 1.12 | | 1.10 | | 1.01 | | 1.01 | |
| Overhead | 1.00 = FP | | | | 1.07 = FP | | | | 1.01 = FP | | | |
| | 1.00 = CA | | | | 1.10 = CA | | | | 1.01 = CA | | | |

**Conclusion**

In this chapter, we presented the simulation results of faults affecting modules of various DMFB architectures; we did the simulations for mixing modules and the splitting/storage/detection modules.

# REFERENCES

[1] D., Grissom, Design of Topologies for Interpreting Assays on Digital Microfluidic Biochips, 2014.

[2] M. N., Gupta, Multi-Board Digital Microfluidic Biochip Synthesis with Droplet Crossover Optimization (Doctoral dissertation, University of Cincinnati), 2014.

[3] R. B., Fair, A., Khlystov, T. D., Tailor, V., Ivanov, R. D., Evans, P. B., Griffin, and J., Zhou, "Chemical and biological applications of digital-microfluidic devices," in *Design & Test of Computers*, 2007.

[4] A., Abdoli, and A., Jahanian, "A General-Purpose Pin-Constrained Digital Microfluidic Biochip," in *18th CSI International Symposium on Computer Architecture and Digital Systems (CADS)*, pp. 1-6, 2015.

[5] A., Abdoli, and A., Jahanian, "Field-Programmable Cell Array Pin-Constrained Digital Microfluidic Biochip," in *22nd Iranian Conference on Biomedical Engineering (ICBME)*, pp. 48-53, 2015.

[6] A., Abdoli, and A., Jahanian, "Fault-Tolerant Architecture and CAD Algorithm for Field-Programmable Pin-Constrained Digital Microfluidic Biochips," in *1st CSI Symposium on Real-Time and Embedded Systems and Technologies (RTEST)*, pp. 1-8, 2015.

[7] M., Yafia, A., Ahmadi, M., Hoorfar, and H., Najjaran, "Ultra-Portable Smartphone Controlled Integrated Digital Microfluidic System in a 3D-Printed Modular Assembly," *Micromachines,* pp. 1289-1305, 2015.

[8] A., Abadian, and S., Jafarabadi-Ashtiani, "Paper-based digital microfluidics," *Microfluidics and Nanofluidics,* pp. 989-995, 2014.

[9] M., Yafia, S., Shukla, and H., Najjaran, "Fabrication of digital microfluidic devices on flexible paper-based and rigid substrates via screen printing," *Journal of Micromechanics and Microengineering,* 2015.

[10] D., Grissom, and P. Brisk, "A field-programmable pin-constrained digital microfluidic biochip," in *Annual Design Automation Conference*, 2013.

[11] F. S., and K. Chakrabarty., "High-level synthesis of digital microfluidic biochips," *ACM Journal on Emerging Technologies in Computing Systems (JETC),* 2008.